\documentclass[11pt,twocolumn]{article}
\usepackage[utf8]{inputenc}
\usepackage[T1]{fontenc}
\usepackage{amsmath}
\usepackage{amsfonts}
\usepackage{amssymb}
\usepackage[version=4]{mhchem}
\usepackage{microtype}
\usepackage{stmaryrd}
\usepackage{graphicx}
\usepackage{microtype}
\usepackage[export]{adjustbox}
\usepackage{hanging}
\graphicspath{ {./images/} }
\usepackage[a4paper, total={6in, 8in}]{geometry}
\title{Polygon Detection from a Set of Lines }
\author{Alfredo Ferreira Jr. Manuel J. Fonseca Joaquim A. Jorge\\
Department of Information Systems and Computer Science,
INESC-ID/IST/University of Lisboa\\
R. Alves Redol, 9, 1000-029 Lisboa, Portugal\\
alfredo.ferreira.jr@inesc-id.pt, mjf@inesc-id.pt, jorgej@acm.org}
\date{October 8, 2003}

\begin{document}
\maketitle

\begin{abstract}
Detecting polygons defined by a set of line segments in a plane is an important step in analyzing vector drawings. This paper presents an approach combining several algorithms to detect basic polygons from arbitrary line segments. The resulting algorithm runs in polynomial time and space, with complexities of $O\left((N+M)^{4}\right)$ and $O\left((N+M)^{2}\right)$ respectively, where $N$ is the number of line segments and $M$ is the number of intersections between line segments. Our choice of algorithms was made to strike a good compromise between efficiency and ease of implementation. The result is a simple and efficient solution to detect polygons from lines.
\end{abstract}

\section*{keywords}Polygon Detection, Segment Intersection, Minimum Cycle Basis

\section{Introduction}
Unlike image processing, where data consist of raster images, our algorithm deals with drawings in vector format, consisting of line segments. This requires completely different approaches, such as described here.

We divide this task into four major steps to perform polygon detection from a set of line segments. First, we detect line segment intersections using the Bentley-Ottmann algorithm~\cite{bentley1979}. The next step creates a graph induced by the drawing, where vertices represent endpoints or proper intersection points of line segments and edges represent maximal relatively open subsegments that contain no vertices. The third step finds the Minimum Cycle Basis (MCB)~\cite{syslo1981} of the graph induced in the previous step, using Horton's algorithm~\cite{horton1987}. The last step constructs a set of polygons based on cycles in the previously found MCB. This is straightforward if we transform each cycle into a polygon, where each node represents a polygon vertex, and each edge in the cycle represents an edge in the polygon.

A previous version of this paper was presented in~\cite{ferreira2003}. In sections \ref{sec:2} and \ref{sec:3}, we describe the four steps of our method. Section \ref{sec:4} presents the whole algorithm and experimental results in section \ref{sec:5}5. Finally, in section \ref{sec:6}, we discuss conclusions and future work. 

\section{Intersection Removal}
\label{sec:2}
In a vector drawing composed of a set of line segments, many intersections might exist between these segments. To detect polygonal shapes, we must remove proper segment intersections, thus creating a new set of line segments in which any pair of segments share at most one endpoint.

\subsection{Finding line segment intersections}
The first step of our approach consists of detecting all $M$ intersections between $N$ line segments in a plane. This is considered one of the fundamental problems of Computational Geometry, and it is known that any algorithm within the algebraic decision tree model has a lower bound of $\Omega(N \log N+M)$ time to solve it ~\cite{benOr1983,clarkson1989}.

In~\cite{balaban1995} Balaban proposes two algorithms for finding intersecting segments, a deterministic and asymptotically optimal for both time $O(N \log N+M)$ and space $O(N)$ algorithm and a simpler one that can perform the same task in $O\left(N \log ^{2} N+M\right)$-time. Before that, Chazelle and Edelsbrunner~\cite{chazelle1992} reached a time optimal algorithm $O(N \log N+M)$ with a space requirement of $O(N+M)$. The randomized approach devised by Clarkson and Shor~\cite{clarkson1989} produced an algorithm for reporting all intersecting pairs that requires $O(N \log N+M)$ time and $O(N)$ space.

In 1979 Bentley and Ottmann proposed an algorithm that solved this problem in $O((N+M) \log N)$ time and $O(N+$ $M)$ space~\cite{bentley1979}. This algorithm is the well-known Bentley-Ottmann algorithm, and after more than 40 years, it is still widely adopted in practical implementations because it is easy to understand and implement~\cite{orourke1998,hobby1999}. In realizing that this is not the most complex part of our approach, we use the Bentley-Ottmann algorithm since its complexity is acceptable for our purposes, and its published implementations are quite simple.

\begin{figure}[t]
\begin{center}
\includegraphics[max width=0.275\textwidth]{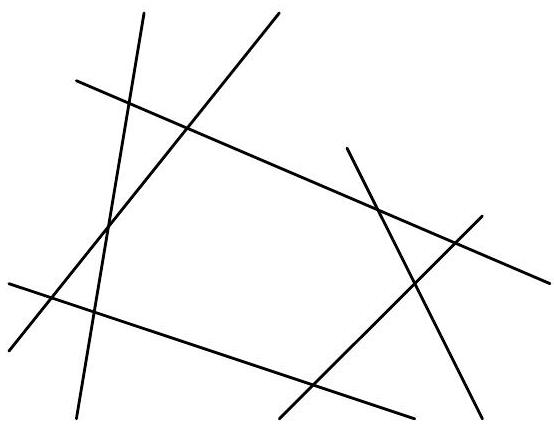}
\end{center}
\caption{Set $\Phi$ of line segments}
\end{figure}

\subsection{Removing line segment intersections}
The next step of our approach is to remove all proper intersections between line segments, dividing each intersected segment in sub-segments without proper intersections, only sharing endpoints. To find and remove intersections, performing at once the first two steps of our approach, we use a robust and efficient implementation of the Bentley-Ottmann algorithm, described by Bartuschka, Mehlhorn and Naher~\cite{bartuschka1997} that computes the planar graph induced by a set of line segments. Their implementation, represented in this paper by \textsc{Compute-Induced-Graph}, computes the graph $G$ induced by set $\Phi$ in $O((N+M) \log N)$ time. Since this algorithm is quite long, we choose not to present it here. We refer our readers to~\cite{bartuschka1997} for a detailed description.

In this implementation, the vertices of $G$ represent all endpoints and proper intersection points of line segments in $\Phi$, and the edges of $G$ are the maximal relatively open sub-segments of lines in $\Phi$ that do not contain any vertex of $G$. The major drawback of this implementation lies in that parallel edges are produced in the graph for overlapping segments. We assume that $\Phi$ contains no such segments. For example, the set $\Phi$ shown in Figure 1, \textsc{Compute-Induced-Graph}, will produce the graph $G$, depicted in Figure 2, where each edge represents a nonintersecting line segment.

\section{Polygon Detection}
\label{sec:3}
Detecting polygons is similar to finding cycles on the graph $G$ produced in the previous step.

The first known linear-time algorithm for listing all graph cycles was presented by Syslo~\cite{syslo1981}. This algorithm requires $O(V)$ space and $O(V \times C)$ time, where $V$ and $C$ are the number of vertices and cycles in $G$, respectively. Later, Dogrusöz and Krishnamoorthy proposed a vector space algorithm for enumerating all cycles of a planar graph that runs in $O\left(V^{2} \times C\right)$ time and $O(V)$ space ~\cite{dogrusoz1995}. Although asymptotically slower, this algorithm is much simpler than Syslo's and is amenable to parallelization. Unfortunately, the number of cycles in a planar graph can grow exponentially with the number of vertices~\cite{mateti1976}. An example of this situation is the graph presented in Figure 3. In this case, the number of cycles, including the interior region numbered 1, is $O\left(2^{r}\right)$ with $r=k / 2+1$, where $k$ is the number of vertices since one can choose any combination of the remaining regions to define a cycle~\cite{dogrusoz1995}. This is why detecting all polygons that can be constructed from a set of lines is not very feasible. In this paper, we chose to detect the minimal polygons with few edges that cannot be constructed by joining other minimal polygons.

\begin{figure}[t]
    \centering
\includegraphics[max width=0.3\textwidth]{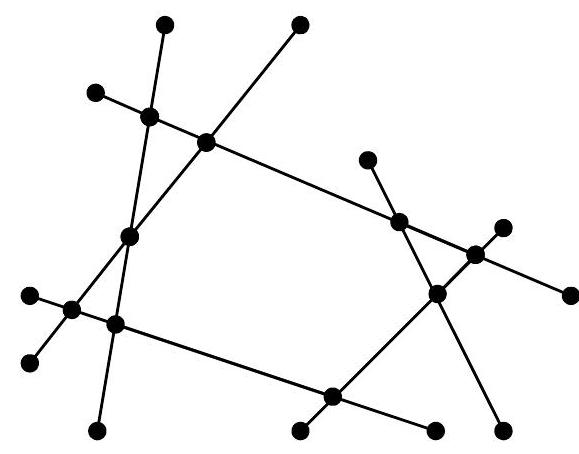}
    \caption{Graph $G$ induced by $\Phi$}
    \label{fig:induced-graph}
\end{figure}

\subsection{Minimum Cycle Basis of a Graph}
Since we want to detect the minimal polygons, this can be treated as searching for a Minimum Cycle Basis (MCB). So, the second step of our approach consists in obtaining an MCB of graph $G$. A cycle basis is defined as a basis for the cycle space of $G$, which consists entirely of elementary cycles. A cycle is called elementary if it contains no vertex more than once. The dimension of the cycle space is given by the cyclomatic number $\nu=E-V+P$~\cite{euler1752,cauchy1813}, where $E$ is the number of edges and $V$ the number of vertices in $G$ and $P$ is the number of connected components of $G$.

\begin{figure}[b]
    \centering
\includegraphics[max width=0.2\textwidth]{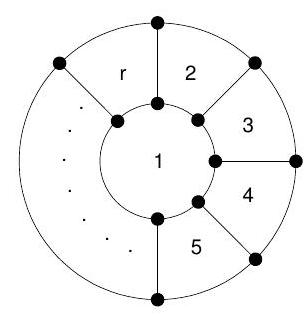}
    \caption{A planar graph with an exponential number of cycles}
    \label{fig:exp-graph}
\end{figure}

\subsection{All Cycles of a Graph}
Horton presented the first known polynomial-time algorithm to find the shortest cycle basis of a graph, which runs in $O\left(E^{3} V\right)$ time~\cite{horton1987} or in $O\left(E^{4}\right)$ on simple planar graphs~\cite{hartvigsen1994}, which is the case. While asymptotically better solutions have been published in the literature, the Bentley-Ottmann algorithm is simple and usable for our needs. The pseudo-code of this algorithm is listed in \textsc{Minimum-Cycle-Basis} and shortly described below. A further detailed description of this algorithm and its concepts can be found in~\cite{horton1987}.

The \textsc{All-Pairs-Shortest-Paths} finds the shortest paths between all pairs of vertices in graph $G$ and can be performed in $O\left(V^{3}\right)$ time and $O\left(V^{2}\right)$ space using FloydWarshall or Dijkstra algorithms~\cite{cormen1990}. \textsc{Order-By-length} orders the cycles by ascending length and can be implemented by any efficient sorting algorithm. This is a non-critical step because it has a $O(V \nu \log V)$ upper bound in time complexity, which is insignificant compared to other steps of this algorithm.

\begin{figure}[t]
    \centering
\includegraphics[max width=0.4\textwidth]{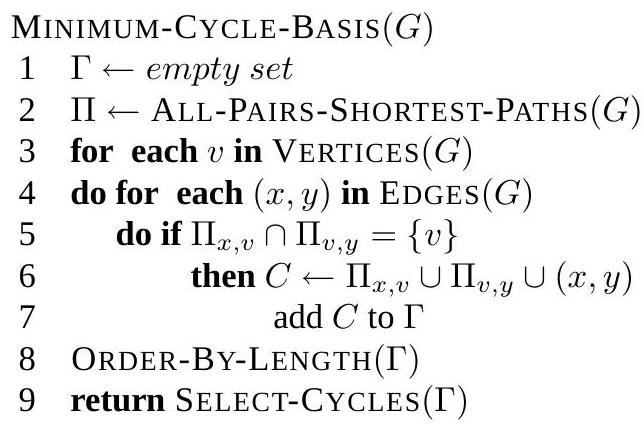}
    \label{fig:enter-label}
\end{figure}

In \textsc{Select-Cycles}, we use a greedy algorithm to find the MCB from $\Gamma$ set of cycles. To do this Horton~\cite{horton1987} suggests representing the cycles as rows of a $0-1$ incidence matrix, in which columns correspond to the edges of the graph, and rows are the incidence vectors of each cycle. Gaussian elimination using elementary row operations over the integers modulo two can then be applied to the incidence matrix, processing each row in turn, in ascending order of the weights of cycles, until enough independent cycles are found.

This step dominates the time complexity from other steps since it takes $O\left(E \nu^{2} V\right)$ time. Knowing that $G$ is always a simple planar graph we can conclude that as a whole, the \textsc{Minimum-Cycle-Basis} algorithm has a worst-case upper bound of $O\left(E \nu^{2} V\right)=O\left(E^{3} V\right)=O\left(E^{4}\right)$ operations and space requirements of $O\left(V^{2}\right)$.

\begin{figure}[b]
    \centering
\includegraphics[max width=0.3\textwidth]{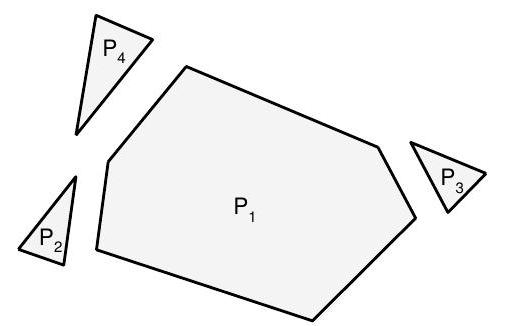}
    \caption{Set $\Theta$ of polygons detected from $\Phi$}
    \label{fig:poly-set}
\end{figure}

Figure~\ref{fig:gamma} shows an example of $\Gamma$, the set of cycles resulting from applying the \textsc{Minimum-Cycle-Basis} to graph $G$ shown in Figure 2.

\subsection{Polygon construction}
The last step of our approach consists of constructing a set $\Theta$ of polygons from the MCB. An algorithm to perform this operation can easily run in $O(C V)$ time, where $C$ is the number of cycles in MCB. Such an algorithm is listed in \textsc{Polygons-From-Cycles}, which returns a set $\Theta$ of polygons.

\begin{figure}[b]
    \centering
\includegraphics[max width=0.3\textwidth]{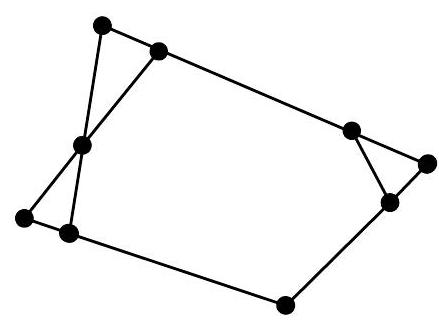}
    \caption{Shortest cycle basis $\Gamma$ of graph $G$}
    \label{fig:gamma}
\end{figure}

\begin{figure}
\begin{center}
\includegraphics[max width=0.4\textwidth]{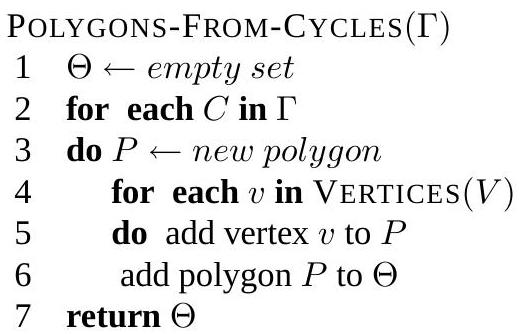}
\end{center}
\end{figure}

Figure~\ref{fig:poly-set} illustrates the resulting set $\Theta$ of polygons generated by applying \textsc{Polygons-From-Cycles} to $\Gamma$ depicted in Figure~\ref{fig:gamma}.

\section{Algorithm Outline}
\label{sec:4}
We can now outline \textsc{Detect-Polygons}. This algorithm can detect a set $\Theta$ of polygons from an initial set $\Psi$ of line segments. To perform this task, we pipeline the algorithms referred to in previous sections for line segment intersection removal, MCB finding, and cycle-to-polygon conversion.

\begin{figure}[b]
    \centering
\includegraphics[max width=0.4\textwidth]{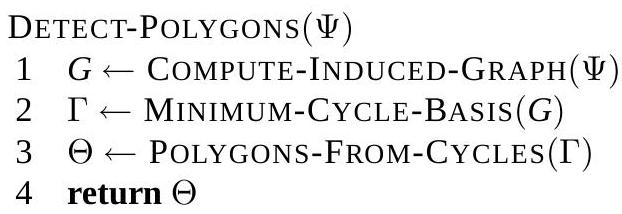}
\end{figure}

As referred in section 2.2, \textsc{Compute-Induced-Graph} runs in $O((N+M) \log N)$ time and $O(N+M)$ space. The \textsc{Shortest-Cycle-Basis} runs in $O\left(V^{4}\right)$ operations and has a space requirement of $O\left(V^{2}\right)$, making this the critical step in the complexity of this algorithm, since the \textsc{Polygons-From-Cycles} needs $O(C V)$ time.

Since the number $V$ of vertices in the graph is no greater than the sum of line endpoints $(2 \times N)$ with detected intersections $M$, we can then conclude that the proposed algorithm has time and space complexities of $O\left(V^{4}\right)=$ $O\left((N+M)^{4}\right)$ and $O\left(V^{2}\right)=O\left((N+M)^{2}\right)$, respectively.

\section{Experimental Results}
\label{sec:5}
The algorithm proposed in this paper was implemented in $\mathrm{C}++$ and tested in an Intel Pentium III 1GHz 512MB RAM computer running Windows XP. We tested the algorithm with line segments created from simple test drawings, technical drawings of mechanical parts, and hand-sketched drawings. Table~\ref{table:results} presents the results obtained from these tests.

These results show that performance is acceptable for online processing in sets with less than three hundred lines, like hand sketches or small-size technical drawings. If the line set has about 2,500 lines, the algorithm will take more than twenty minutes to detect the polygons. Still, this remains a feasible solution for batch processing of medium-size technical drawings.

\section{Conclusions and Future Work}
\label{sec:6}
The proposed algorithm uses polygon detection in vector drawings to create descriptions based on spatial and topological relationships between polygons. Another use is detecting planar shapes in sketches. Both applications have been implemented as working prototypes for shape retrieval and architectural drawing from sketches.

The algorithm presented here detects all minimal polygons that can be constructed from a set of line segments in polynomial time and space. This approach uses well-known and simple-to-implement algorithms to perform line segment intersection detection and to find an MCB of a graph instead of using more efficient but less simple methods.

Indeed, the presented algorithm has considerable room for improvement, namely through more recent, complex, and efficient algorithms. Further work may be carried out regarding detecting and correcting rounding errors resulting from finite precision computations.

\section*{Acknowledgments}
We thank Professor Mukkai S. Krishnamoorthy from Rensselaer Polytechnic Institute, New York, for his very helpful suggestions.

This work was partly funded by the Portuguese Foundation for Science and Technology, project 34672/99, and the European Commission, project SmartSketches IST-200028169.

\begin{table}[t]
\scalebox{0.89}{
\begin{tabular}{|r|r|r|r|r|}
\hline
Lines & Intersections & Nodes & Edges & Time (ms) \\
\hline\hline
6 & 9 & 21 & 24 & 10 \\
\hline
36 & 16 & 58 & 68 & 50 \\
\hline
167 & 9 & 169 & 177 & 3986 \\
\hline
286 & 47 & 389 & 376 & 8623 \\
\hline
518 & 85 & 697 & 679 & 36703 \\
\hline
872 & 94 & 1066 & 10050 & 128995 \\
\hline
2507 & 10 & 2407 & 2526 & 1333547 \\
\hline
\end{tabular}
}
\caption{Results of algorithm tests}
\label{table:results}
\end{table}

\nocite{Fonseca2005,Ferreira2009,pereira2004}
\bibliographystyle{plain}
\small
\bibliography{references} 

\begin{thebibliography}{10}

\bibitem{balaban1995}
Ivan~J. Balaban.
\newblock An optimal algorithm for finding segment intersections.
\newblock In {\em Proceedings of the 11th Annual ACM Symposium Computational Geometry}, pages 211--219. ACM, 1995.

\bibitem{bartuschka1997}
Ulrike Bartuschka, Kurt Mehlhorn, and Stefan Naher.
\newblock A robust and efficient implementation of a sweep line algorithm for the straight line segment intersection.
\newblock In {\em Proceedings of Workshop on Algorithm Engineering}, pages 124--135, Venice, Italy, September 1997.

\bibitem{benOr1983}
M.~Ben-Or.
\newblock Lower bounds for algebraic computation trees.
\newblock In {\em Proceedings of the 15th Annual ACM Symposium Theory of Computing}, pages 80--86. ACM, 1983.

\bibitem{cauchy1813}
Augustin-Louis Cauchy.
\newblock Recherche sur les polyèdres.
\newblock {\em J. Ecole Polytechnique}, 9(16):68--86, 1813.

\bibitem{chazelle1992}
Bernard Chazelle and Herbert Edelsbrunner.
\newblock An optimal algorithm for intersecting line segments in the plane.
\newblock {\em Journal of the ACM}, 39:1--54, 1992.

\bibitem{clarkson1989}
Ken Clarkson and Peter~W. Shor.
\newblock Applications of random sampling in computational geometry, ii.
\newblock {\em Discrete and Computational Geometry}, 4:387--421, 1989.

\bibitem{cormen1990}
Thomas Cormen, Charles Leiserson, and Ronald Rivest.
\newblock {\em Introduction to Algorithms}.
\newblock MIT Press, McGraw-Hill, 2nd edition, 1990.

\bibitem{dogrusoz1995}
U.~Dogrusöz and M.~Krishnamoorthy.
\newblock Cycle vector space algorithms for enumerating all cycles of a planar graph.
\newblock Technical Report~5, Rensselaer Polytechnic Institute, Dept. of Computer Science, Troy, New York 12180 USA, January 1995.

\bibitem{euler1752}
Leonhard Euler.
\newblock Elementa doctrinae solidorum.
\newblock {\em Novi Commentarii Academiae Scientiarum Petropolitanae}, 4:109--140, 1752.

\bibitem{Ferreira2009}
Alfredo Ferreira, Simone Marini, Marco Attene, Manuel~J. Fonseca, Michela Spagnuolo, Joaquim~A. Jorge, and Bianca Falcidieno.
\newblock Thesaurus-based 3d object retrieval with part-in-whole matching.
\newblock {\em International Journal of Computer Vision}, 89(2–3):327–347, June 2009.

\bibitem{Fonseca2005}
Manuel~J. Fonseca, Alfredo Ferreira, and Joaquim~A. Jorge.
\newblock Content-based retrieval of technical drawings.
\newblock {\em International Journal of Computer Applications in Technology}, 23(2/3/4):86, 2005.

\bibitem{hartvigsen1994}
David Hartvigsen and Russel Mardon.
\newblock The all-pairs minimum cut problem and the minimum cycle basis problem on planar graphs.
\newblock {\em SIAM Journal on Computing}, 7(3):403--418, August 1994.

\bibitem{hobby1999}
John Hobby.
\newblock Practical segment intersection with finite precision output.
\newblock {\em Computational Geometry: Theory and Applications}, 13(4), 1999.

\bibitem{horton1987}
J.D.Horton.
\newblock A polynomial-time algorithm to find the shortest cycle basis of a graph.
\newblock {\em SIAM Journal on Computing}, 16(2):358--366, April 1987.

\bibitem{bentley1979}
J.L.Bentley and T.Ottmann.
\newblock Algorithms for reporting and counting geometric intersections.
\newblock {\em IEEE Transactions on Computers}, pages 643--647, 1979.

\bibitem{ferreira2003}
Alfredo~Ferreira Jr., Manuel Fonseca, and Joaquim Jorge.
\newblock {Polygon Detection from a Set of Lines}.
\newblock In Adérito Marcos, Ana Mendonça, Miguel Leitão, António Costa, and Joaquim Jorge, editors, {\em 12º Encontro Português de Computação Gráfica}. The Eurographics Association, 2003.

\bibitem{mateti1976}
Prabhaker Mateti and Narsingh Deo.
\newblock On algorithms for enumerating all circuits of a graph.
\newblock {\em SIAM Journal on Computing}, 5(1):90--99, March 1976.

\bibitem{orourke1998}
Joseph O'Rourke.
\newblock {\em Computational Geometry in C}, chapter Section 7.7 "Intersection of Segments", pages 264--266.
\newblock Cambridge University Press, 2nd edition, 1998.

\bibitem{pereira2004}
Nelson~F Silva.
\newblock Cascading recognizers for ambiguous calligraphic interaction.
\newblock In {\em EUROGRAPHICS Workshop on Sketch-Based Interfaces and Modeling}. The Eurographics Association, 2004.

\bibitem{syslo1981}
Maciej~M. Syslo.
\newblock An efficient cycle vector space algorithm for listing all cycles of a planar graph.
\newblock {\em SIAM Journal on Computing}, 10(4):797--808, November 1981.

\end{thebibliography}

\end{document}